\documentclass[%
 reprint,
 amsmath,amssymb,
 aps,
pra,
]{revtex4-2}

\usepackage[dvipsnames, svgnames, x11names]{xcolor}
\usepackage{graphicx}
\usepackage{epstopdf}
\usepackage{dcolumn}
\usepackage{bm}
\usepackage{amsmath}
\usepackage{multirow}
\usepackage{booktabs}
\usepackage{amsfonts}
\usepackage{makecell}
\usepackage{float}
\usepackage{diagbox}
\usepackage{tabularx}
\usepackage{tabu}                        
\usepackage{array}
\usepackage{amssymb}

\begin{document}

\preprint{APS/123-QED}

\title{High-tolerance antiblockade SWAP gates using optimal pulse drivings }

\author{Wan-Xia Li$^{1}$} 
\author{Jin-Lei Wu$^{2} $}
\author{Shi-Lei Su$^{2}$} 
\author{Jing Qian$^{1,3}$} 

\affiliation{$^{1}$State Key Laboratory of Precision Spectroscopy, Department of Physics, School of Physics and Electronic Science, East China Normal University, Shanghai, 200062, China
}
\affiliation{$^{2}$School of Physics, Zhengzhou University, Zhengzhou 450001, China}
\affiliation{$^{3}$Shanghai Branch, Hefei National Laboratory, Shanghai 201315, China}



\begin{abstract}
Position error is treated as the leading obstacle  that prevents Rydberg antiblockade gates from being experimentally realizable, because of the inevitable fluctuations in the relative motion between two atoms invalidating the antiblockade condition. In this work we report progress towards a high-tolerance antiblockade-based Rydberg SWAP gate enabled by the use of {\it modified} antiblockade condition combined with carefully-optimized laser pulses.
Depending on the optimization of diverse pulse shapes our protocol shows that the amount of time-spent in the double Rydberg state can be shortened by more than $70\%$ \textcolor{black}{with respect to the case using {\it perfect} antiblockade condition}, which significantly reduces this position error. Moreover, we benchmark the robustness of the gate via taking account of the technical noises, such as the Doppler dephasing due to atomic thermal motion, 
the fluctuations in laser intensity and laser phase and the intensity inhomogeneity.
As compared to other existing antiblockade-gate schemes the predicted gate fidelity is able to maintain at above 0.91 after a very conservative estimation of various experimental imperfections, especially considered for realistic interaction deviation of $\delta V/V\approx 5.92\%$ at $T\sim20$ $\mu$K. 
Our work paves the way to the experimental demonstration of Rydberg antiblockade gates in the near future.
\end{abstract}


\email{jqian1982@gmail.com}
\maketitle
\preprint{}
\section{introduction}

Implementation of Rydberg antiblockade gates is one of the current mainstream protocols towards fast and robust neutral-atom quantum computation \textcolor{black}{\cite{saffman2016quantum,henriet2020quantum,morgado2021quantum,cong2022hardware,shi2022quantum,graham2022multi}}, as the traditional Rydberg antiblockade regime with enhanced dissipative dynamics can lead to the preparation of high-fidelity steady entanglement in versatile systems \textcolor{black}{\cite{carr2013preparation,shao2014stationary,su2015simplified,barontini2015deterministic,chen2017dissipative,2017Robust,chen2018accelerated,wintermantel2020unitary,PhysRevLett.125.133602,PhysRevA.103.023335,PhysRevLett.128.080503}}.
Beyond applications in quantum entanglement,
constructing a quantum logic gate via antiblockade effect \textcolor{black}{\cite{PhysRevLett.98.023002,amthor2010evidence,bai2020distinct}} is more straightforward because the simultaneous multi-atom excitation establishes an effective model which exactly avoids the occupancy of lossy singly-excited Rydberg states. Despite these significant advances,
however, owing to the leading participation of multi-atom excited states, such antiblockade gates are found to be extremely sensitive to the relative position fluctuations between two atoms \textcolor{black}{\cite{saffman2005analysis,li2013nonadiabatic,shi2017annulled,robicheaux2021photon}} as well as to the atomic decays \textcolor{black}{\cite{de2018analysis}}, which inherently forbids its  experimental validation yet.

Based on antiblockade mechanism, some prior studies of e.g.
control-phase gates \textcolor{black}{\cite{su2017applications}}, presented the gate infidelity far larger than 0.15 for an interaction deviation of $\delta V/V\approx 5.0\%$. More recently,
Wu {\it et.al.} \textcolor{black}{\cite{wu2021one}} demonstrated two-atom SWAP gates with an infidelity of $\sim 0.03$ when the position deviation in the interatomic distance is only $\sim3.0$ nm, necessarily for using a deep cooling of ground-state qubits \textcolor{black}{\cite{PhysRevX.2.041014,hu2017creation,PhysRevResearch.5.033093}} or sufficiently deep traps
\textcolor{black}{\cite{yang2016coherence}}. Therefore, although the antiblockade mechanism can lead to high-fidelity Rydberg gates in theory, to implement them in experiment remains untouchable yet. A realistic position-error tolerant scheme is urgently needed.

In this paper we present a practical scheme to realize the antiblockade SWAP gates with greatly improved tolerance, especially to the significant fluctuations in interatomic interaction. We 
optimize different laser pulse shapes for minimizing the time spent in the double Rydberg state, so as to decrease the sensitivity of gate respect to the interatomic distance deviation. It is observable that, as accompanied by a {\it modified} antiblockade condition $V = \Delta_1+\Delta_2+q$ which contains a nonzero factor $q$ to loosen the antiblockade constraint, the occupancy of double Rydberg state can be strongly suppressed 
for an arbitrary pulse shape \textcolor{black}{\cite{pagano2022error}}. \textcolor{black}{Remember in a {\it perfect} antiblockade-gate scheme where the interaction energy is perfectly compensated by the detunings between the driving field and atomic transition, i.e. $V = \Delta_1+\Delta_2$, 
it is challenging to construct quantum gate operations since some unwanted ac Stark shift terms would emerge in the effective Hamiltonian(see Eq.\ref{eff}), although the excitation of single atom has been eliminated by off-resonant couplings.}
We find that the ideal two-qubit SWAP gates in the absence of any position deviation, have an average fidelity of $\mathcal{F}\sim 0.9997$. In addition even taking account of other technical imperfections a more conservative estimation arises {$\mathcal{F}\geq 0.9140$} for a typical $1.0$-$\mu$s gate duration, which is mainly caused by the position error $\sim 0.0752$ considered for an atomic temperature of $T \sim 20$ $\mu$K (equivalently $\delta V/V \approx 5.92$ $\%$). 
\textcolor{black}{We also verify that the characteristic time scale for the gate is determined by the maximal Rabi frequency of the driving lasers. When their maxima are restricted to be smaller than $2\pi\times 50$ MHz cooperating with the optimization for the gate duration, a faster($\sim 0.1177$ $\mu$s) antiblockade SWAP gate is achievable which contains a higher tolerance to the position error due to the minimal time spent, see an extended study in appendix \ref{ab}.}

Our implementation building on the earlier proposal of antiblockade gates \textcolor{black}{\cite{wu2021one}}, additionally adopts optimal time-dependent pulse drivings, which, not only avoids the combination of a series of elementary gates \cite{PhysRevA.101.062312} and realizes the nontrivial SWAP gate within one-step implementation; but also 
greatly improves the antiblockade-gate robustness against fundamental position fluctuations between two trapped atoms. The SWAP gate is important with extensive applications in e.g., quantum entanglement swapping \textcolor{black}{\cite{riebe2008deterministic,ma2012experimental,ning2019deterministic}} and quantum repeater \textcolor{black}{\cite{sangouard2011quantum,PhysRevA.106.062611,PhysRevA.108.022609}}. As compared to other existing proposals the parameters we set to realize the gate, are all extracted from careful optimization which can provide an opportunity to decrease the leading obstacle i.e., the position error that mainly restricts the antiblockade gate fidelity.


\section{Theoretical strategy}

\begin{figure}
\centering
\includegraphics[width=8.65cm,height=4.3cm]{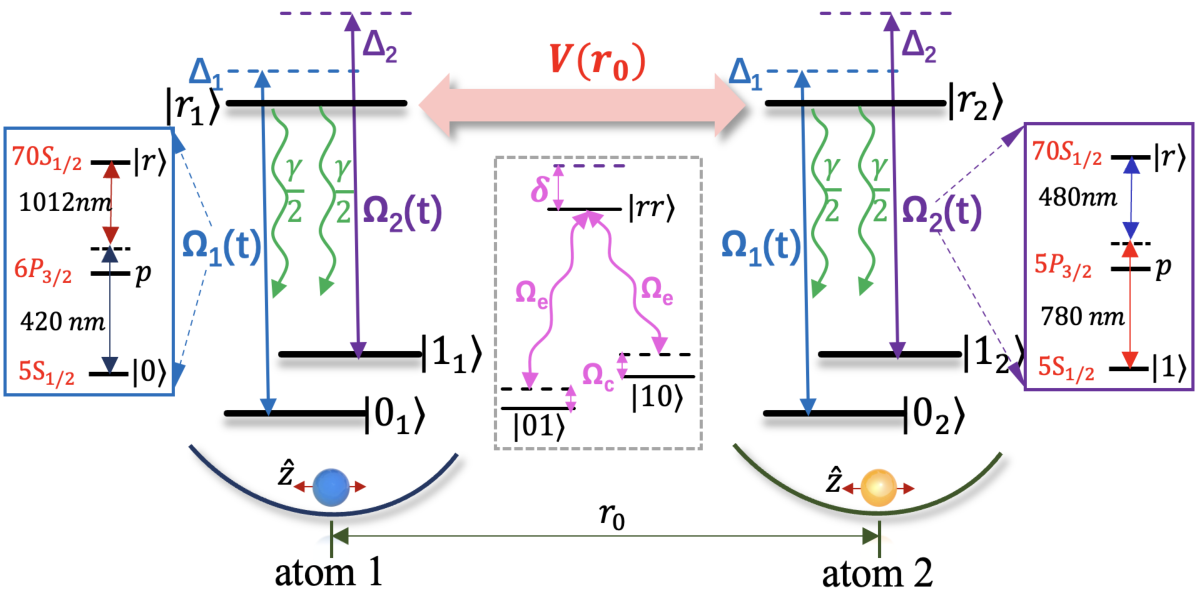}
\caption{Level structure of two-qubit Rydberg SWAP gates based on {\it modified} antiblockade. The central inset gives an effective swapping model formed by an off-resonance Raman-like transition on two targeted states $|01\rangle$ and $|10\rangle$ mediated by $|rr\rangle$. The atomic qubits are defined as hyperfine clock qubits encoded in ground energy levels $|0_{1,2}\rangle=|5S_{1/2},F=1,m_F=0\rangle$ and $|1_{1,2}\rangle=|5S_{1/2},F=2,m_F=0\rangle$. 
$|r_{1,2}\rangle=|70S_{1/2}\rangle$ is a Rydberg state considered for a rubidium-87 experiment \textcolor{black}{\cite{PhysRevLett.123.170503}}. The {\it vdWs} dispersion coefficient for the doubly-excited state $|70S_{1/2};70S_{1/2}\rangle$ is $C_6/2\pi=862.69$ GHz$\cdot \mu$m$^6$ \textcolor{black}{\cite{vsibalic2017arc}}, leading to the non-fluctuated Rydberg interaction strength $V/2\pi\approx 70.49$ MHz at a fixed distance $r_{0} = 4.8$ $\mu$m. The Rydberg decay rate is $\gamma = 2\pi/\tau$ with $\tau(\approx 400$ $\mu$s) being the lifetime of Rydberg states. See text for more details.}
\label{Fig1:model}
\end{figure}

In this work we propose a two-qubit SWAP gate with high position-error tolerance based on {\it modified} antiblockade condition.
As illustrated in Fig.\ref{Fig1:model}, two identical neutral atoms individually trapped in two optical tweezers, have hyperfine ground states $|0\rangle$, $|1\rangle$ and Rydberg state $|r\rangle$. When atoms are simultaneously excited they interact via the Rydberg-Rydberg interaction. Here the interaction strength is quantified by $V = C_6/r_0^6$ denoting a van der Waals ({\it vdWs})-type interaction.
During the laser-based operation the excitation $|0\rangle\to|r\rangle$ in each atom is accomplished by a two-photon process with effective Rabi frequency $\Omega_1(t)$ and two-photon detuning $\Delta_1$. Similarly the other transition $|1\rangle\to|r\rangle$ is accomplished by $\Omega_2(t)$ and $\Delta_2$. Such two-photon processes are usually mediated via different low-lying $p$ states (Fig.\ref{Fig1:model} insets) which have been adiabatically eliminated due to the large intermediate detunings for $p$ states \textcolor{black}{\cite{PhysRevA.99.043404}}. 
In addition we assume state $|r\rangle$ decays to $|0\rangle$, $|1\rangle$ with rate $\gamma$ and equal branching ratios. \textcolor{black}{To be more realistic we will introduce a leakage state $|\alpha\rangle$(not shown in Fig.\ref{Fig1:model}) for each atom outside of the qubit basis $\{|0\rangle,|1\rangle\}$. Here we restrict the analysis to the case of $\eta_{r\to 0(1)} = 1/2$, $\eta_{r\to \alpha} = 0$ which provides an ideal estimation on the gate fidelity.}

With the rotating-wave approximation the total Hamiltonian for this system can be described by ($\hbar=1$ hereafter)
\begin{equation}
\hat{\mathcal{H}}=\hat{\mathcal{H}}_{1}\otimes \hat{\mathcal{I}}_2+\hat{\mathcal{I}}_1 \otimes\hat{ \mathcal{H}}_{2} + \hat{V}_{rr}
\label{Ham}
\end{equation}
where
\begin{equation}
\hat{\mathcal{H}_{j}}=\Omega_1e^{-i\Delta_1 t}|0\rangle_j\langle r|+\Omega_2e^{-i\Delta_2 t}|1\rangle_j\langle r|+\text{H.c}
\end{equation}
accounts for the $j$th atom coupling to the driving lasers
and
\begin{equation}
 \hat{V}_{rr} = V |rr\rangle\langle rr|  
\end{equation}
is the interaction between the Rydberg states. Here $j\in (1,2)$ and $\hat{\mathcal{I}}_j$ denotes the identity matrix.

While deriving the effective form of $\hat{\mathcal{H}}$ through the second-order perturbation calculation we apply two parameter conditions. One is $|\Delta_1|\gg\Omega_1^{\max}$, $|\Delta_2|\gg\Omega_2^{\max}$ which aims at canceling the population on singly-excited Rydberg states by off-resonant couplings. The resulting single excitation terms are all decoupled to the initial states and can be discarded, see Eq.(\ref{eff00}).
\textcolor{black}{The other is $V = \Delta_1+\Delta_2+q $ representing a {\it modified} antiblockade condition in which the presence of factor $q$ is crucial for suppressing unwanted population on state $|rr\rangle$. Notice that 
$q=0$ reduces to the case of {\it perfect} antiblockade in which the double Rydberg state $|rr\rangle$ can obtain a facilitated excitation because of the perfect compensation between
$V$ and $\Delta_1+\Delta_2$ \cite{PhysRevLett.128.013603}.
But, when $q\neq 0$ known as the {\it modified} antiblockade one can modify the antiblockade regime by changing the Stark shift terms of $|rr\rangle$ and therefore shorten the time spent on it \cite{su2016one}.}

After rotating with respect to the operator $\hat{U}=e^{-i(\Delta_1+\Delta_2)t|rr\rangle\langle rr|}$ the effective formula of $\hat{\mathcal{H}}$ can be separately written as
 \begin{eqnarray}
       \hat{\mathcal{H}}_{00,eff} = \frac{2\Omega_1^2}{\Delta_1}|00\rangle\langle 00| \label{00e} \\ \hat{\mathcal{H}}_{11,eff} = \frac{2\Omega_2^2}{\Delta_2}|11\rangle\langle 11| \label{11e}
   \end{eqnarray}
for initial states $|00\rangle$ and $|11\rangle$,
\begin{equation}
\begin{aligned}
  \hat{\mathcal{H}}_{01,eff}&= \Omega_e(|rr \rangle\langle 01|+ |rr \rangle\langle 10| )  +\text{H.c.}\\ &+\Omega_c(|01 \rangle\langle 01|+|10 \rangle\langle 10|)+\delta|rr \rangle\langle rr|  
   \label{eff}
   \end{aligned}
\end{equation}
for initial states $|01\rangle$ where
\begin{equation}
  \Omega_e =\frac{\Omega_1\Omega_2}{\Delta_1}+\frac{\Omega_1\Omega_2}{\Delta_2},
  \Omega_c =\frac{\Omega_{1}^2}{\Delta_1}+\frac{\Omega_{2}^2}{\Delta_2},
  \delta = \frac{2\Omega_{2}^2 }{\Delta _{1}}+
   \frac{2\Omega_{1}^2 }{\Delta _{2}}+q
\end{equation}
respectively denote the effective couplings, the ac Stark shift with respect to $|01\rangle$ and $|10\rangle$ and the effective detuning(including ac Stark shifts) to $|rr\rangle$. From Eqs.(\ref{00e}-\ref{eff}) we see states $|00\rangle$ and $|11\rangle$ are dark to all gate pulses as expected.
Whereas if the initially-considered states are $|01\rangle$ or $|10\rangle$
the system will reduce to an effective three-level swapping model.
Since states $|10\rangle$ and $|01\rangle$ exhibit equivalent dynamics in this symmetric setup so $\hat{\mathcal{H}}_{10,eff}=\hat{\mathcal{H}}_{01,eff}$.
More details about the derivation of Eqs.(\ref{00e}-\ref{eff}) can be found in \textcolor{black}{appendix \ref{aa}}.

From $\hat{\mathcal{H}}_{01,eff}$
we find an effective three-level system for states $|01\rangle$ and $|10\rangle$ which is off-resonantly coupled to $|rr\rangle$ (Fig.\ref{Fig1:model} inset).
Thus one can expect the combined action of all quantities including $\Omega_e$, $\Omega_c$, $\delta$, will achieve a perfect state swapping by obeying
\begin{equation}
 |01\rangle \rightleftarrows |rr\rangle  \rightleftarrows |10\rangle.  
\end{equation}
However, 
the participation of simultaneous Rydberg excitation state $|rr\rangle$ in this process will no doubt suffer from a serious position error, mainly caused by the significant deviation of interatomic interaction $\delta V$(with respect to the non-fluctuated strength $V$) between individually trapped atoms that breaks the antiblockade condition \textcolor{black}{\cite{PhysRevA.90.032329}}. 
Fortunately, the effective detuning $\delta$ to state $|rr\rangle$ contains a factor $q$ which provides a more flexible adjustment to the original ac Stark shift term $\frac{2\Omega_{2}^2 }{\Delta _{1}}+
   \frac{2\Omega_{1}^2 }{\Delta _{2}}$ on $|rr\rangle$. We show a positive $q$ that modifies the ac Stark shift, 
   can make state $|rr\rangle$ far off-resonance so as to reduce the population on it \textcolor{black}{\cite{PhysRevLett.129.200501}}. Therefore our protocol can exhibit a higher tolerance against significant interaction fluctuations arising from e.g. the thermal motion of atoms in the inhomogeneous coupling field \cite{PhysRevResearch.5.033052} or laser intensity noise \textcolor{black}{\cite{jiang2023sensitivity}}.

\section{Gate implementation with optimal pulses}

\begin{figure*}
\centering
\includegraphics[width=17.5cm,height=10.5cm]{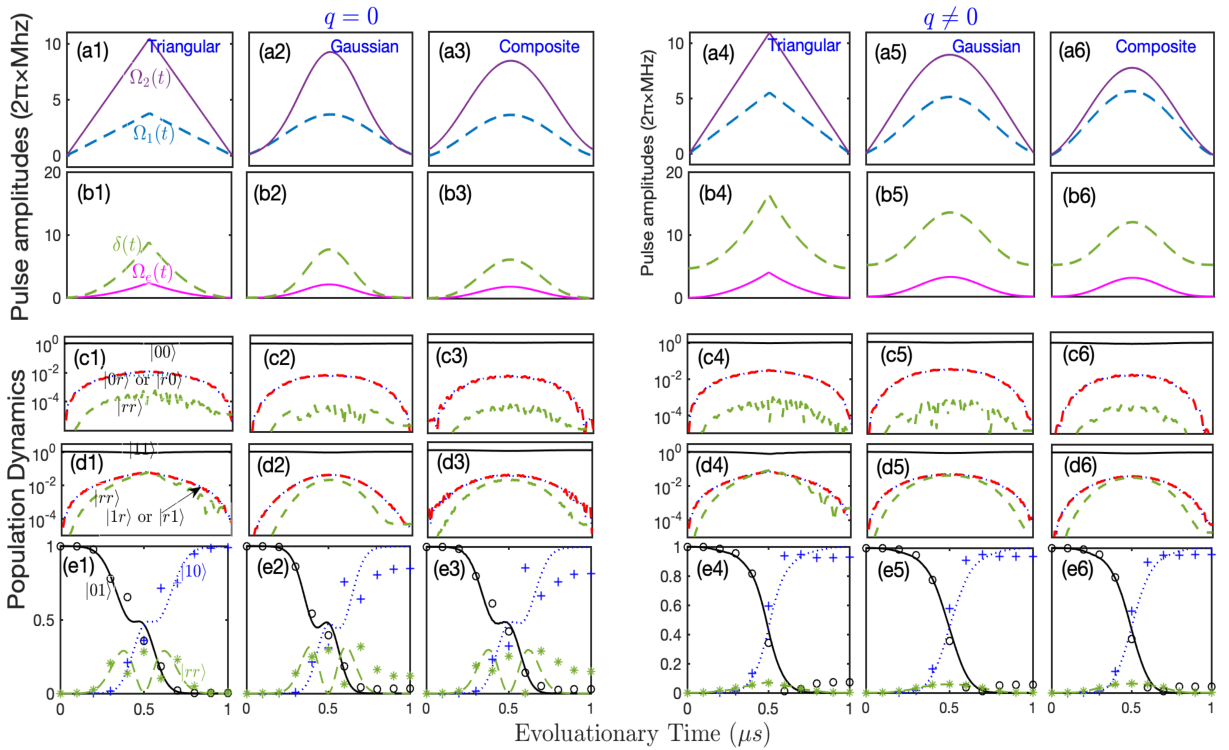}
\caption{Realization of two-qubit SWAP gates with different optimal pulse shapes.
(a1,a4) Optimal isosceles-triangle shaped Rabi frequencies $\Omega_1(t)$, $\Omega_2(t)$ and (b1,b4) optimal effective coupling $\Omega_e(t)$ and effective detuning $\delta(t)$ {\it versus} the evolutionary time $t$ where $T_g=1.0$ $\mu$s. $q=0$ and $q\neq 0$ stand for the {\it perfect} antiblockde and {\it modified} antiblockade cases. The corresponding time-dependent population dynamics based on the original Hamiltonian $\hat{\mathcal{H}}$ for different initial states $\{|00\rangle,|11\rangle,|01\rangle\}$, are  respectively displayed in (c1,c4), (d1,d4), (e1,e4).
In (e1) and (e4) the population dynamics of $|01\rangle$(black-circles), $|10\rangle$(blue-crosses), $|rr\rangle$(green-stars) calculated by the effective Hamiltonian $\hat{\mathcal{H}}_{01,eff}$ are also comparably displayed. Similar results with optimal Gaussian and composite pulse shapes are given in 
(a2-e2; a5-e5) and (a3-e3; a6-e6). }
\label{Fig2:dyna}
\end{figure*}

Now we show that the two-qubit SWAP gates can be realized for diverse pulse shapes $\Omega_1(t),\Omega_2(t)$. We perform global numerical optimization for all pulse parameters and laser detunings by using the genetic algorithm as in our prior work \textcolor{black}{\cite{Li_2023}}, which aims at maximizing the ideal gate fidelity. \textcolor{black}{Notice that all stochastic deviations arising from technical imperfections are excluded in the perfect optimization case, and the gate robustness against main error sources will be analyzed after the perfect optimization (sections \ref{IV} and \ref{V}).
A different choice of optimization algorithm combined with a careful consideration for the gate robustness to some type of errors \cite{PRXQuantum.4.020336}, will leave for our future work.}

To fully understand the population dynamics we numerically solve the two-atom master equation in the Lindblad form \textcolor{black}{\cite{PhysRevA.95.012708}}, 
\begin{equation}
    \dot{{\rho}} = -i[{\hat{\mathcal{H}}},\hat{\rho}] +\frac{1}{2}\sum_{j=1}^6[2\hat{\mathcal{L}}_j\hat{\rho}\mathcal{L}_j^\dagger -(\hat{\mathcal{L}}_j^\dagger\hat{\mathcal{L}}_j\hat{\rho}+\hat{\rho}\hat{\mathcal{L}}_j^\dagger\hat{\mathcal{L}}_j)]
    \label{rhoh}
\end{equation}
where the single-atom decay operator $\hat{\mathcal{L}}_j$ expressed in the basis $\{|0\rangle,|1\rangle,|r\rangle,|\alpha\rangle\}$($|\alpha\rangle$ is a leakage state) is given by
\begin{eqnarray}
\hat{\mathcal{L}}_{1}&=&\sqrt{\eta_{r\to 0}\gamma}|0\rangle_1\langle r|,\hat{\mathcal{L}}_{2}=\sqrt{\eta_{r\to 0}\gamma}|0\rangle_2\langle r| \nonumber\\
\hat{\mathcal{L}}_{3}&=&\sqrt{\eta_{r\to 1}\gamma}|1\rangle_1\langle r|,
\hat{\mathcal{L}}_{4}=\sqrt{\eta_{r\to 1}\gamma}|1\rangle_2\langle r| \\
\hat{\mathcal{L}}_{5}&=&\sqrt{\eta_{r\to \alpha}\gamma}|\alpha\rangle_1\langle r|,
\hat{\mathcal{L}}_{6}=\sqrt{\eta_{r\to \alpha}\gamma}|\alpha\rangle_2\langle r| \nonumber
\end{eqnarray}
representing all Rydberg-state scattering channels and $|r\rangle\to|\alpha\rangle$ presents the leakage out of the qubit manifold to other $m_F$ states. \textcolor{black}{In the calculation we have assumed state $|r\rangle$ decays to the ground-state manifold $|0\rangle$ and $|1\rangle$ uniformly with equal probability $\eta_{r\to 0}=\eta_{r\to 1}=1/2$, and $\eta_{r\to \alpha}=0$.
An extended calculation based on realistic eight ground-state manifold will be discussed in appendix \ref{ab}.}
Numerical solutions of Eq.(\ref{rhoh}) can be directly used to extract the average gate fidelity in the perfect optimization case over four input states $\{|00\rangle,|01\rangle,|10\rangle,|11\rangle\}$ as
\begin{equation}
\mathcal{F}=\frac{1}{4} \text{Tr}\sqrt{\sqrt{U} \rho(t=T_{g})\sqrt{U}}
\label{fde}
\end{equation}
with ``4" the number of input states, 
$U$ the ideal transfer matrix for the SWAP operation and $\rho(t)$ the practical density matrix measured at $t=T_g$ (gate duration). For any input state \textcolor{black}{one can obtain full population dynamics of arbitrary intermediate state} along with the above master equation (\ref{rhoh}).

To obtain an antiblockade SWAP gate with high tolerance the leading step is to find out optimal pulse shapes with an appropriate $q$, where the time spent in the Rydberg level can be minimized, especially in the double Rydberg state $|rr\rangle$. 
For comparison we demonstrate the gate using different pulse shapes as done by \textcolor{black}{\cite{PhysRevA.94.032306}}. \textcolor{black}{The characteristic time scale $T_g$ for the gate is constrained by the maximal Rabi frequency $(\Omega_1^{\max},\Omega_2^{\max})$ of the driving lasers that should meet the swapping circles. Since here we focus on comparing the gate performance under different pulse shapes, $T_g$ is fixed to $1.0$ $\mu$s for simplicity. This choice restricts the maximal strength of Rabi frequencies to be about $\sim2\pi\times 10$ MHz. 
A larger Rabi frequency($\sim 2\pi\times 50$ MHz) can further decrease the $T_g$ value to the level of $\sim 0.1$ $\mu$s, promising for the realization of a faster gate with improved robustness. The relationship between $T_g$ and $\Omega_{1,2}^{\max}$ is numerically confirmed by 
an extended study involving an optimal $T_g$
(appendix \ref{ab}).}
Besides, in order to show the significance of $q$ we seek for optimal pulse parameters individually under two cases: $q=0$ and $q\neq 0$ which are summarized in Table I. The corresponding full population dynamics are comparably displayed in Fig. \ref{Fig2:dyna}. 

\begin{table}
\caption{\label{tab:table1} Optimal pulse parameters and optimal laser detunings with different pulse shapes for $q=0$ and $q\neq 0$. Parameters such as $\Omega_{j}^{\max}$, $\Omega_{0j}$, $\Omega_{1j}$, $\Omega_{2j}$, $\Delta_1$, $q$ are all in units of ($2\pi\times$MHz), and $\omega_j$ is in unit of ($\mu$s). In any case the antiblockade condition $V = \Delta_1 + \Delta_2 + q$ is always satisfied and $V=2\pi\times70.49$ MHz treats as the ideal(non-fluctuated) interaction strength between two atoms at a fixed distance $r_0$. The gate duration $T_g=1.0$ $\mu$s is unvaried.
The ideal gate fidelity $\mathcal{F}$ calculated in the absence of any technical imperfection is given in the last column. The Rydberg decay error intrinsically caused by the limited lifetime of Rydberg levels, can be roughly estimated by $1-\mathcal{F}$.}
\renewcommand{\arraystretch}{1.8}
\setlength{\tabcolsep}{0.8mm}
   \centering
  \resizebox{\linewidth}{!}{ 
   \begin{tabular}{l c c c c}
        \hline
        \hline
         \multicolumn{5}{c}{{\it Case i}: Isosceles-triangle shape} \\
        \hline
    $q$ & $\Omega_{1}^{\max}$ & $\Omega_{2}^{\max}$ & $ \Delta_1$& $\mathcal{F}$ \\
        \hline
        $0$ & 3.769 & 10.441 &  26.642& 0.9992  \\
        \hline
        $4.668$ & 5.513 & 10.947 &  23.090& 0.9996  \\
        \hline
 \multicolumn{5}{c}{{\it Case ii}: Gaussian shape}\\
        \hline
    $q$ & $(\Omega_{1}^{\max},\Omega_{2}^{\max})$& $ (\omega_1,\omega_2)$& $\Delta_1$&$\mathcal{F}$ \\ 
        \hline
        $0$ & (3.595,9.219) & (0.279,0.189)&23.924&0.9993\\ 
        \hline
        $5.037$ & (5.114,8.948)& (0.257,0.351)&22.110&0.9997\\ 
        \hline
 \multicolumn{5}{c}{{\it Case iii}: Composite shape}\\
        \hline       $q$ & $(\Omega_{01},\Omega_{11},\Omega_{21})$&$(\Omega_{02},\Omega_{12},\Omega_{22})$&$\Delta_1$&$\mathcal{F}$ \\ 
        \hline
        $0$ &(1.033,-1.167,1.334)&(3.100,-2.662,2.617)&25.194&0.9992 \\
        \hline
        $5.090$ &(1.774,-1.884,2.036)&(2.558,-2.624,2.608)&22.532&0.9997  \\
        \hline
        \hline        
 \end{tabular}
}
\end{table}

First we demonstrate the gate with two optimal isosceles-triangle shaped pulses via assuming
\begin{eqnarray}
  \text{{\it Case i}:    } \Omega_j(t) &=&  \Omega_j^{\max} \text{               when $t=\frac{T_g}{2}$}   \nonumber \\
  \Omega_j(t) &=&  0 \text{               
                 when $t=0,T_g$}   \nonumber
\end{eqnarray}
This triangular pulse shape only involves one parameter $\Omega_j^{\max}$ to be optimized except $\Delta_1$ and $q$.
Seen from Fig.\ref{Fig2:dyna}(a1-e1) and (a4-e4), we observe that, although the peak Rabi frequencies $\Omega_{1}^{\max}$ and $\Omega_{2}^{\max}$ are comparable with each other; yet case (a4-e4) with $q\neq 0$ clearly causes a large energy shift for the effective detuning $\delta(t)$(green-dashed line in (b4)) which enables a big reduction in the time spent of double Rydberg state. We verify this by calculating the duration to be in state $|rr\rangle$ for input $|01\rangle$ and find that in (e1) with $q=0$, the time spent defined by a time-integration is
$T_{rr}=\int_0^{T_g}\rho_{rr}(t)dt\approx 0.0926$ $\mu$s. While this value can reduce to $T_{rr}\approx0.0248$ $\mu$s in the case of (e4) where $q/2\pi=4.668$ MHz(optimal), which shortens the time spent in the double Rydberg state by $73.2\%$.
Such a large reduction directly provides an improvement in the average gate fidelity yielding $\mathcal{F}= 0.9996$(remember $\mathcal{F}= 0.9992$ for $q=0$) due to the decrease of the Rydberg decay loss in the perfect case. In addition we illustrate the validity of the effective Hamiltonian (\ref{eff}) by simulating the effective population dynamics of states  $|01\rangle$, $|rr\rangle$ and $|10\rangle$ in (e1) and (e4). A good coincidence between the full and effective models are explicit confirming the accuracy of our perturbation theory \textcolor{black}{\cite{wu2020effective}}.

Furthermore, by taking account of the convenience of experimental demonstration, we turn to realize the SWAP gate with smooth-amplitude pulses which contain more variables to be optimized \textcolor{black}{\cite{PhysRevA.105.042430}}. \textcolor{black}{On the experimental side, the smooth-amplitude gate can help to reduce off-resonant coupling to other states which are not included in the idealized model description and provide a finite bandwith of the laser pulses reducing the experimental complexity on fast switching of them \cite{evered2023high}}. Here two smooth pulse shapes are considered which are
\begin{eqnarray}
  \text{{\it Case ii}:    }  \Omega_j(t) &=& \Omega_{j}^{\max}[\frac{e^{{-{(t-{T_g}/{2})}^2}/{2{\omega_j}^2}}-e^{{-{({T_g}/{2})}^2}/{2{\omega_j}^2}}}{1-e^{{-{({T_g}/{2})}^2}/{2{\omega_j}^2}}} ]\nonumber\\
   \text{{\it Case iii}:    } \Omega_j(t) &=& \Omega_{0j} + \Omega_{1j}\cos(\frac{2\pi t}{T_g})+\Omega_{2j}\sin(\frac{\pi t}{T_g}). \nonumber
\end{eqnarray}
while keeping 
$\Delta_{1}$ and $q$ optimal values(Table I). $\Delta_2 =V - \Delta_1-q$ is always fulfilled.
Particularly, in {\it Case ii} the Gaussian function has been corrected by allowing an exact zero intensity at the start and end of the pulse, which, from an experimental point of view, has a better feasibility \textcolor{black}{\cite{PhysRevApplied.18.044042}}.
Finally we present the optimal smooth-amplitude pulses and the population dynamics separately in Fig.\ref{Fig2:dyna}(a2-e2;a5-e5) and (a3-e3;a6-e6). It is explicit that, when $q\neq 0$ the time spent in state $|rr\rangle$ remains at a low level, i.e. $T_{rr}\approx (0.0238,0.0254)$ $\mu$s, arising a slightly higher fidelity $\mathcal{F}\approx 0.9997$ than the triangular pulse case. Therefore, although the effective state swapping between $|01\rangle$ and $|10\rangle$ also requires the participation of $|rr\rangle$, its influence has been largely suppressed by applying pulse optimization under the {\it modified} antiblockade condition.

\textcolor{black}{By now we have presented the realization of high-fidelity SWAP gates with an arbitrary shape for laser pulses. The key in pulse optimization lies in minimizing the time spent in $|rr\rangle$ by letting $\delta(t)$ off-resonance in the presence of $q$ which leads to a direct coupling between $|01\rangle$ and $|10\rangle$. Based on the effective three-level model (Fig.\ref{Fig1:model}) we
evaluate the coupling strength by adiabatically eliminating $|rr\rangle$ and find the effective pulse area calculated by $ \int_0^{T_g}\frac{2\Omega_e(t)^2}{\delta(t)}\approx (3.2600,3.2318,3.1946)$ are all close to $\pi$ for three cases (i-iii) which strongly confirms the importance of optimizing pulse area (not pulse shape) in the realization of two-qubit SWAP gates. The resulting gate infidelity mainly caused by the Rydberg decay denoted as $1-\mathcal{F}= (3\sim4)\times10^{-4}$, is no doubt very robust to arbitrary pulse shapes.
}


The participation of antiblockade condition can facilitate simultaneous excitation from the ground state to the double Rydberg state, strongly suppressing the influence from intermediate singly-excited states, which really gives rise to an efficient one-step implementation of Rydberg gates. Nevertheless, 
the leading obstacle that limits the robustness of such antiblockade gates,  becomes
its sensitivity in practice to the interaction fluctuations between two Rydberg-state atoms. Typically they do depend on ``frozen" Rydberg atomic interactions so as to fulfill the antiblockade condition \textcolor{black}{\cite{PhysRevLett.100.253001,PhysRevLett.110.133001}}. In the present work by minimizing the time spent in $|rr\rangle$($\sim 0.02$ $\mu$s) together with the merit of small Rydberg decay errors ($\sim10^{-4}$), the gate tolerance to the interaction fluctuations can be largely improved, promising for a high-fidelity quantum gate with hotter atoms.

\section{Position Error tolerance} \label{IV}

\begin{figure*}
\centering
\includegraphics[width=11.7cm,height=7.5cm]{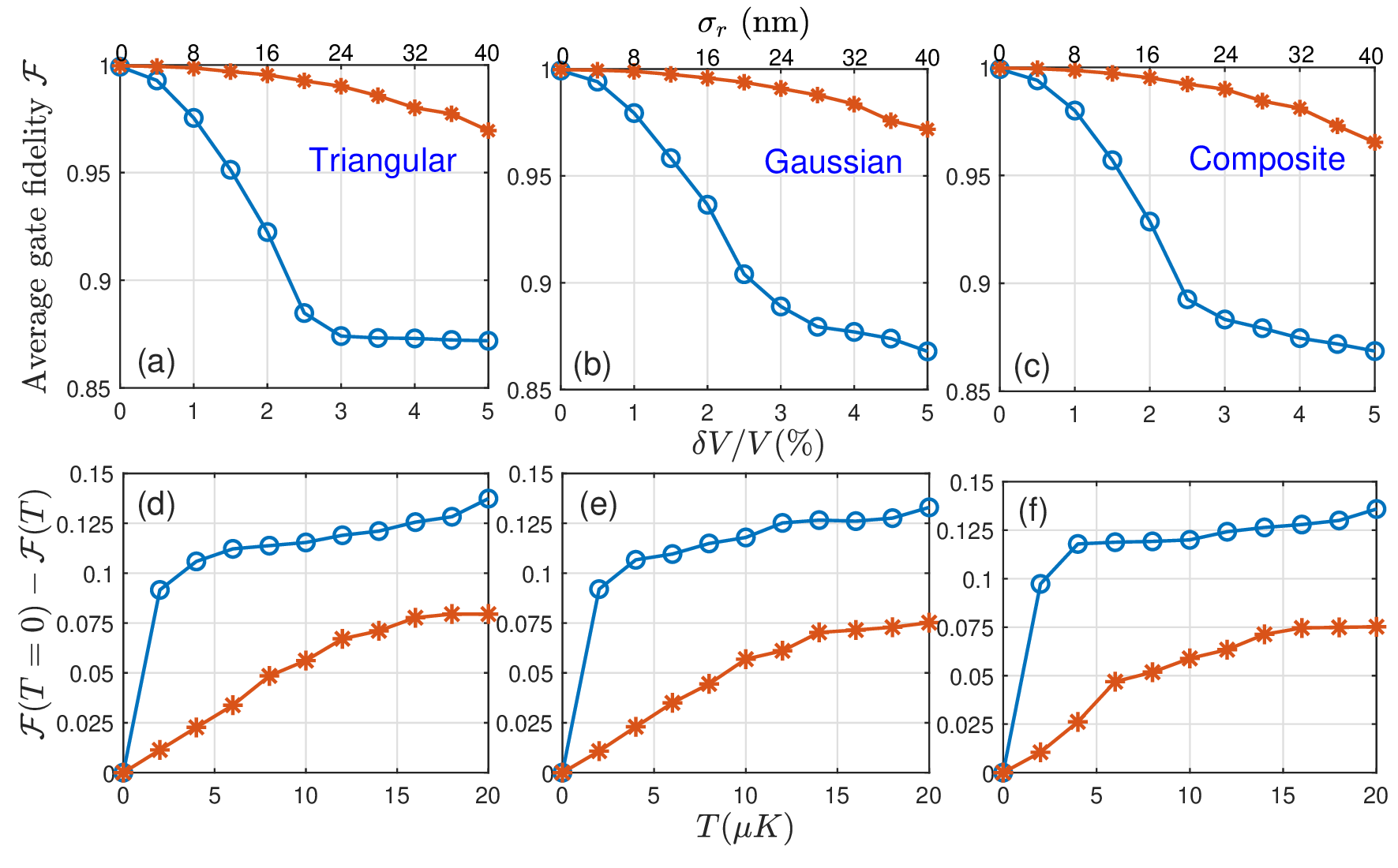}
\caption{(a-c) Average gate fidelity $\mathcal{F}$ as a function of the deviation degree of interaction strength $\delta V/V\in [0,0.05]$(equivalent to the position deviation $\sigma_r\in[0,40]$ nm), where the decay from the Rydberg state is considered. (d-f) Gate infidelity denoted by $\mathcal{F}(T=0)-\mathcal{F}(T)$ under different atomic temperatures $T$, which is treated as one of the dominant error sources to produce an interaction deviation. 
Cases $q=0$ and $q\neq 0$ are presented by the blue-line with circles and the red-line with stars, respectively.
\textcolor{black}{Each point in (a-c) denotes an average of 300 uniformly-distributed random samplings. While in (d-f) each point denotes an average of 300 Gaussian-distributed samplings for the atomic position. Pulse parameters used in the calculation are all obtained in the perfect case without any technical noise (Table I)}.}
\label{Fig3:perturb}
\end{figure*}

It is well known that, a Rydberg antiblockade gate depends on the leading participation of double Rydberg states, and is therefore more sensitive to the fluctuation of interatomic interaction than other gate schemes, such as Rydberg blockade \textcolor{black}{\cite{PhysRevLett.104.010503,PhysRevLett.104.010502}} or adiabatic passage \textcolor{black}{\cite{PhysRevA.89.030301,PhysRevA.94.062307,PhysRevA.96.042306,PhysRevA.98.052324,PhysRevA.97.032701,PhysRevA.101.062309,PhysRevA.101.030301}}. 
In reality this fluctuation in the relative position between two atoms can not be avoided, especially for two trapped atoms under a finite temperature \textcolor{black}{\cite{hutzler2017eliminating}}. That is the reason why the Rydberg antiblockade gates have not been reported by experiments yet. Although, e.g. placing two atoms at a long distance \textcolor{black}{\cite{,PhysRevA.95.052330,shi2021quantum}} or lower the atomic temperature \textcolor{black}{\cite{kaufman2021quantum}} can reduce the position error the robustness of antiblockade gates is still unsatisfactory.

In this section, in order to explore the influence of interatomic position fluctuation, we first assume a relative deviation ratio $\delta V/V$ with respect to the non-fluctuated interaction strength $V$ by taking a replacement of
\begin{equation}
    V \to V(1 + \delta V/V)
\end{equation}
where $\delta V/V$ stands for the maximal deviation degree of the position fluctuation. \textcolor{black}{For a determined $\delta V/V$ we extract a uniformly-distributed random number generated from the range of $[-\delta V/V,+\delta V/V]$ which
leads to a fluctuated interaction strength.}
Fig.\ref{Fig3:perturb}(a-c) plot the average fidelity $\mathcal{F}$ {\it versus} the variation of $\delta V/V\in[0,0.05]$. Such deviation degree can also be changed into the view of two-atom distance fluctuation because $\sigma_r= (\delta V/V)r_0/6$ due to $V= C_6/r_0^6$ for a {\it vdWs} interaction \cite{PhysRevLett.118.063606}.
From Figs.\ref{Fig3:perturb}(a-c) it is apparent that, with an increasing deviation the SWAP gate fidelity significantly decreases for the $q=0$ case. E.g. when $\delta V/V = 0.05$ (equivalent to a position deviation of $\sigma_r\approx40$ nm), $\mathcal{F}$ is decreased to $\sim0.87$ with arbitrary pulse shapes. However, with the help of {\it modified} antiblockade condition where $q$ is optimized to minimize the time spent in $|rr\rangle$, the gate robustness against the deviation degree $\delta V/V$ can be greatly improved. The average gate fidelity $\mathcal{F}$ stays as high as $\sim0.97$ at a large distance deviation.

Accounting for the fact that deviation in the Rydberg interaction between two atoms usually comes from the imperfections of atomic cooling and trapping, so we consider the atoms with a finite temperature $T$ and re-calculate this position error.
We assume that the relative position between two atoms satisfies a one-dimensional Gaussian distribution with its mean $r_0=(C_6/V)^{1/6}\approx4.8$ $\mu$m and standard deviation $\sigma_{r} = \sqrt{k_BT/m\omega^2}$ which varies with $T$. Here, $k_B$, $m$, $\omega$ mean the Boltzmann constant, the atomic mass and the trapping frequency. 
By assuming $\omega/2\pi=147$ kHz, $T=20$ $\mu$K we can obtain  $\sigma_r \approx 47.34$ nm and $\delta V/V \approx 5.92\% $. From the numerical results in Fig.\ref{Fig3:perturb}(d-f), it clear that, the gate infidelity $\mathcal{F}(T=0)-\mathcal{F}(T)$(exclude the decay error) for the {\it modified} antiblockade case is dramatically decreased. At $T= 20$ $\mu$K accessible by a typical cold-atom experiment \textcolor{black}{\cite{PhysRevA.97.063423}}, we find that the gate infidelity due to the position error can sustain $\sim 7.52\times 10^{-2}$, which is almost a 50$\%$ reduction as compared to that for the $q=0$ case. \textcolor{black}{A further expectation for decreasing the position error down to the level of $\sim 10^{-3}$ may resort to lowering the temperature $T$ or increasing the trapping frequency $\omega$ which allows for a smaller position deviation $\sigma_r< 14.7$ nm. Then the influence of position fluctuation to the infidelity can be suppressed to an experimentally accessible level $\sim 9.6\times 10^{-3}$.}
Among the existing schemes of Rydberg antiblockade gates \textcolor{black}{\cite{shao2014one,su2021dipole}}, our protocol reports the gate with the best position-error tolerance making one-step closer to its experimental demonstration in the near future.

\section{Robustness against technical imperfections} \label{V}

In this section we study the gate infidelity by taking account of other technical imperfections, as considered by an existing Rydberg experiment \textcolor{black}{\cite{PhysRevLett.123.230501}}. We will restrict our analysis to the experimentally-accessible Gaussian pulse ({\it Case ii}) which possesses a more smooth adjustment of the Rabi frequency as well as the best non-fluctuated (ideal) gate fidelity in the perfect case. Except for the position error having explored in Sec. \ref{IV}, other technical obstacles to a higher gate fidelity may involve the Doppler effect due to the atoms having a certain speed, the intensity and phase noises in excitation laser pulses, the relative position of atoms within a finite laser spatial width {\it et.al}. \textcolor{black}{\cite{zhang2012fidelity}}. In the following we analyze in detail the influence of each contribution to the gate infidelity in order to give a conservative estimation for the gate performance.

\begin{figure}
\centering
\includegraphics[width=8.5cm,height=7.5cm]{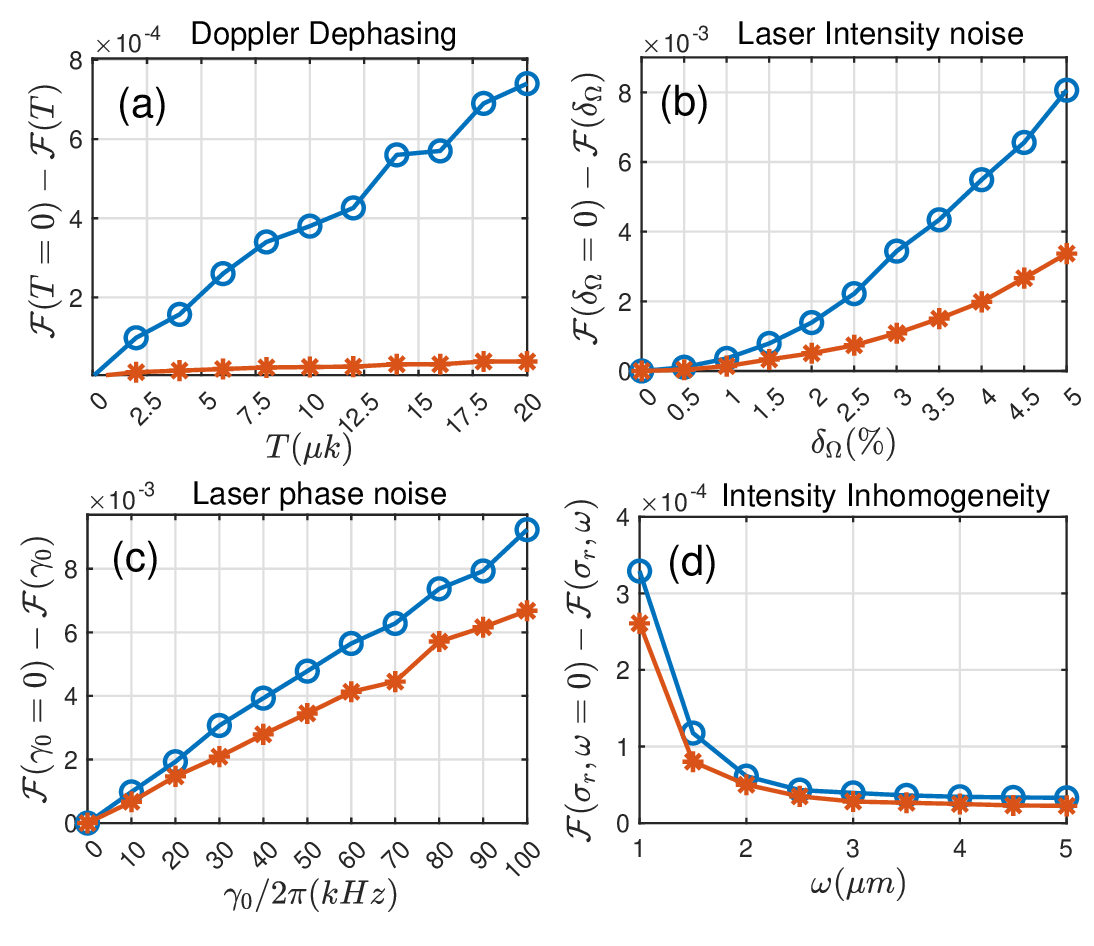}
\caption{Gate infidelity of the antiblockade SWAP gate with a pair of optimal Gaussian shaped pulses({\it Case ii}), caused by (a) Doppler dephasing under different atomic temperatures $T\in[0,20]$ $\mu$K, (b) Laser intensity fluctuations $\delta_{\Omega}\in[0,0.05]$, (c) Laser phase fluctuations(characterized by a dephasing rate $\gamma_0/2\pi\in[0,100]$ kHz for the ground-Rydberg transition), and (d) Laser beam waist $\omega\in[1,5]$ $\mu$m where the standard deviation of the atomic position is fixed to $\sigma_r \approx 47.34$ nm at $T=20$ $\mu$K. Similar to Fig.\ref{Fig3:perturb}, numerical results based on the $q=0$ case(blue-line with circles) and the $q\neq 0$ case(red-line with stars) are  respectively presented.
Each point is obtained by an average of 300 samplings. \textcolor{black}{Note that all simulation results depend on the choice of optimal pulse parameters in the perfect {\it Case ii} (Table I).}}
\label{Fig4:errors}
\end{figure} 

{\it Doppler dephasing.-} In a practical environment the qubit atoms would have a certain speed due to the residual thermal motion arising the Doppler effect. The resulting laser frequency detuning felt by the atoms will be slightly different from its desired value $\bar{\alpha}=0$, which gives rise to a modified one-atom Hamiltonian
\begin{equation}
\hat{\mathcal{H}_{j}}=\Omega_1e^{-i(\Delta_1+\alpha_1) t}|0_j\rangle\langle r_j|+\Omega_2e^{-i(\Delta_2+\alpha_2) t}|1_j\rangle\langle r_j|+\text{H.c}
\end{equation}
where $\alpha_{1(2)} = \vec{k}_{1(2),eff}\cdot\vec{v}$ stands for an extra phase factor to the Rabi frequency and should fulfill an one-dimensional Gaussian distribution
\begin{equation}
    f(\alpha_{1,2}) = \frac{1}{\sqrt{2\pi}\sigma_{\alpha_{1,2}}}e^{-\frac{(\alpha_{1,2}-\bar{\alpha})^2}{2\sigma_{\alpha_{1,2}}^2}}
    \label{gau}
\end{equation}
with respect to the standard deviations $\sigma_{\alpha_{1}} = k_{1,eff}v_{rms} $ and $\sigma_{\alpha_{2}} = k_{2,eff}v_{rms} $. Given $T=20$ $\mu$K corresponds to the one-dimensional rms velocity $v_{rms} = \sqrt{k_BT/m}\approx 44$ mm/s. In addition,
we recall our setup as shown in Fig.\ref{Fig1:model}(insets) where both transitions $|0\rangle\to|r\rangle$ and $|1\rangle\to|r\rangle$ are mediated by two different low-lying $p$ states. Therefore the effective wavevectors are ${k}_{1,eff} = 8.76\times 10^6$ m$^{-1}$, ${k}_{2,eff} = 5\times 10^6$ m$^{-1}$ when two excitation lasers have a counter-propagating configuration to minimize thermal Doppler shifts. That means for each measurement the extra frequency detuning $\alpha_{1,2}$(except $\Delta_{1,2}$) of excitation lasers seen by the atoms is a random value extracted from a centered Gaussian function 
$f(\alpha_{1,2})$ with maximal standard deviations $(\sigma_{\alpha_{1}},\sigma_{\alpha_{2}}) = (0.383,0.219) $ MHz. Numerical simulation allows for a new estimation of the gate infidelity $\mathcal{F}(T=0)-\mathcal{F}(T)$, shown in Fig.\ref{Fig4:errors}(a) {\it versus} the atomic temperature $T$. By averaging over sufficient runs of measurement, the gate infidelity based on {\it modified} antiblockade mechanism explicitly shows a dramatic reduction by one order of magnitude as compared to that for the {\it perfect} antiblockade case. The reason is that $\alpha_{1(2)}$ serves as an extra and fluctuated detuning in addition to $\Delta_{1(2)}$, and the presence of an optimal $q$ no doubt adds to the tolerance against the detuning deviation. We find that, at $T=20$ $\mu$K the infidelity caused by the Doppler dephasing has been suppressed to a negligible level $\sim 3.80\times 10^{-5}$ for the $q\neq 0$ case.

{\it Laser intensity noise.-} In general the laser intensity fluctuation coming from external technical weakness, will lead to the deviation of desired excitation probability. Especially a gate scheme that depends on pulse optimization, requires an accurate knowledge of pulse parameters \cite{PhysRevApplied.13.024059}. Here we introduce a same relative deviation $\delta_{\Omega} = \delta\Omega/\Omega^{\max}_{1,2}$ in the range of $\delta_{\Omega}\in[0,0.05]$ with respect to the maximal Rabi frequency $\Omega_{1,2}^{\max}$. During each run of measurement the realistic laser Rabi frequency $\Omega_{1(2)}(t)$ should be fluctuated by a uniform random number obtained from $[-\delta_{\Omega},+\delta_{\Omega}]\Omega^{\max}_{1(2)} $ while the pulse shape is kept unchanged.
Fig.\ref{Fig4:errors}(b) shows the dependence of the gate infidelity on the laser intensity fluctuation. By increasing the relative deviation $\delta_{\Omega}$ from 0 to 5.0 $\%$, the {\it modified} antiblockade case apparently has a better robustness to the laser intensity noise, although 
the gate infidelities preserve at the level of $\sim 10^{-3}$ in
both cases.

{\it Laser phase noise.-} The phases $\phi_{1,2}(t)$ to the laser Rabi frequency $\Omega_{1,2}(t)$ are random processes which should be characterized by a phase noise spectral density as a function of the Fourier frequency \textcolor{black}{\cite{di2010simple}}. Since the laser phase noise caused by different frequencies involved in excitation lasers, would lead to dephasing of Rabi oscillations \textcolor{black}{\cite{PhysRevLett.121.123603,madjarov2020high,PhysRevApplied.15.054020}}, we describe average result of the laser phase noise by introducing an extra Lindblad superoperator $\hat{\mathcal{L}}_d[\rho]$ in the master equation (\ref{rhoh})
\begin{equation}
   \hat{\mathcal{L}}_d[\rho] = \frac{1}{2}\sum_{j=1}^4[2\hat{\mathcal{L}}_{dj}\hat{\rho}\mathcal{L}_{dj}^\dagger -(\hat{\mathcal{L}}_{dj}^\dagger\hat{\mathcal{L}}_{dj}\hat{\rho}+\hat{\rho}\hat{\mathcal{L}}_{dj}^\dagger\hat{\mathcal{L}}_{dj})]
    \label{rho}
\end{equation}
with four dephasing channels expressed as
\begin{eqnarray}
\hat{\mathcal{L}}_{d,1(2)}&=&\sqrt{\frac{\gamma_{0}}{2}}(|r\rangle_{1(2)}\langle r|-|0\rangle_{1(2)}\langle 0|) \nonumber\\ 
\hat{\mathcal{L}}_{d,3(4)}&=&\sqrt{\frac{\gamma_{0}}{2}}(|r\rangle_{1(2)}\langle r|-|1\rangle_{1(2)}\langle 1|) \nonumber
\end{eqnarray}
and $\gamma_0$ denotes a common dephasing rate. We have neglected the dephasing channel between $|r\rangle$ and $|\alpha\rangle$ since no laser pulse couples them.
Fig.\ref{Fig4:errors}(c) predicts the relationship between gate infidelity and different dephasing rates where $\gamma_0/2\pi\in[0,0.1]$ MHz is considered. \textcolor{black}{Here we set the maximal $\gamma_0$ to be $2\pi\times 0.1$ MHz at a same level to an experimental value \cite{PhysRevA.101.043421}.}
For each $\gamma_0$ we adopt a uniform random value in the range of $[0,\gamma_0]$ and calculate the average infidelity $\mathcal{F}(\gamma_0=0)-\mathcal{F}(\gamma_0)$ after running 300 samplings. 
We observe that the effect of phase noise is more influential than that of the intensity noise, and the {\it modified} antiblockade condition can also slightly improve the gate tolerance. When $\gamma_0/2\pi=0.1$ MHz the gate infidelity can achieve $6.68\times 10^{-3}$ which also plays a non-negligible role as compared to the dominant position error($\sim 10^{-2}$).

{\it Inhomogeneity in Rabi frequencies.-}
In the above discussions we treat the thermal fluctuations of interatomic position in one dimension which is also the propagation direction for two excitation lasers, e.g., see Fig.\ref{Fig1:model} along $\hat{z}$. In fact since lasers also have finite beam waists, such thermal fluctuations will simultaneously make the atoms deviate from the laser center in the $\hat{x}-\hat{y}$ plane. As a result the actual laser intensity seen by the atoms would deviate from its ideal value, arising a position-dependent Rabi frequency only in the $\hat{x}-\hat{y}$ plane. As a consequence the original time-dependent Rabi frequencies $\Omega_1(t)$ and $\Omega_2(t)$ should be modified by adding a space-dependent factor,
\begin{eqnarray}
    \Omega_{1,2}(t,x,y) = \Omega_{1,2}(t,0)e^{-\frac{x^2}{{w_{x}}^2}-\frac{y^2}{{w_{y}}^2}}
\end{eqnarray}
where $\Omega_{1,2}(t,0)=\Omega_{1,2}(t)$ treats as the central Rabi frequency, and $\omega_{x}=\omega_{y}=\omega$ is assumed to be a common beam waist. We have ignored the intensity inhomogeneity in $\hat{z}$ direction because the position deviation $\sigma_r$ of atoms is usually smaller than the Rayleigh length by orders of magnitude \textcolor{black}{\cite{zhang2012fidelity}}. Furthermore position $(x,y)$ of two atomic qubits in the $\hat{x}-\hat{y}$ plane is modeled by a two-dimension Gaussian function
with standard deviation $\sigma = \sigma_x=\sigma_y\approx 47.34$ nm at $T=20$ $\mu$K. In Fig.\ref{Fig4:errors}(d) we tune the beam waist $\omega$ in the range of $[1,5]$ $\mu$m to see the effect of intensity inhomogeneity. Both cases clearly lead to similar gate infidelity values due to their comparable laser intensities. When $\omega > 2.0$ $\mu$m the influence can be lowered to $< 10^{-4}$. However if $\omega$ is narrowed to be below 1.0 $\mu$m we find the infidelity has a dramatic increase because of the large intensity variations felt by the atoms when the beam waist is too small. While accounting for the fact that the excitation laser can be focused to a wide waist of $\sim 10$ $\mu$m typically \textcolor{black}{\cite{PhysRevLett.100.113003}}, this error can be totally negligible in the scheme.

\section{Discussion and Conclusion}

Ever since the pioneering works Refs. \textcolor{black}{\cite{PhysRevLett.85.2208,PhysRevLett.87.037901}} proposed the idea of quantum computation with neutral atoms a large number of groups have focused on the experimental demonstration of various Rydberg quantum gates via Rydberg blockade mechanism \textcolor{black}{\cite{urban2009observation,gaetan2009observation,PhysRevLett.119.160502}}. This method serves as the leading mainstream blueprint mainly because of its powerful insensitivity to the variation of interatomic interactions that fundamentally makes the position error negligible. In addition, another set of universal quantum gates based on Rydberg antiblockade effect also emerges, since the first scheme by Carr and Saffman proposed a new method for preparing Rydberg entangled states via the combination of dissipative dynamics and antiblockade \textcolor{black}{\cite{carr2013preparation}}. Nevertheless, such antiblockade gates require an exact knowledge of the interaction strength $V$, and thus are very sensitive to the interaction deviation $\delta V$ in practice. In this work we restrict our attention to the implementation of a two-qubit SWAP gate based on antiblockade mechanism, and show that it is possible to improve the tolerance of gates to the position error caused by e.g. residual thermal motion of atoms in the trapping region.

\begin{table}
\caption{\label{tab:table2} Error budget for {\it Case ii} with optimal Gaussian pulses. As expected, among all obstacles(intrinsic and technical) the position error originating from fluctuations in the relative position between two atoms, is the dominant error source.
The Doppler dephasing and the inhomogeneity in laser Rabi frequencies are relatively less important. Every error number is obtained via an average over 300 repeated measurements and the values in parenthesis denote the maximal deviation.
By considering all error sources simultaneously we obtain a very conservative estimate for the gate fidelity, which are $\mathcal{F} \geq 0.8482$ and $\mathcal{F} \geq 0.9143$, corresponding to the cases of $q=0$ and $q\neq 0$ respectively.}
\renewcommand{\arraystretch}{1.6}
\setlength{\tabcolsep}{0.8mm}
   \centering
  \resizebox{\linewidth}{!}{ 
   \begin{tabular}{c c c}
        \hline
        \hline
    Error Sources & \multicolumn{2}{c}{Error Budget} \\
        \hline
         $q/2\pi$ & 0 & 5.037 MHz   \\
        \hline
        Rydberg decay & $7.31\times10^{-4}$ & $3.18\times10^{-4}$  \\
       \hline 
  Position fluctuation~($\sigma_{r}=47.34$ nm) & $1.33\times 10^{-1}$& $7.52\times 10^{-2}$\\
  Doppler dephasing~($T=20$ $\mu$K) &$7.39\times10^{-4}$&$3.80\times10^{-5}$\\
  Inhomogeneous Rabi frequency~($\omega=2$ $\mu$m)& $6.13\times10^{-5}$&$5.02\times10^{-5}$\\
   \hline
  Laser intensity~($\delta_{\Omega}=5.0$ $\%$)&$8.06\times10^{-3}$&
  $3.37\times10^{-3}$\\
Laser phase~($\gamma_0/2\pi=0.1$ MHz)&$9.23\times10^{-3}$&
  $6.68\times10^{-3}$\\
        \hline
        \hline        
\end{tabular}}
\end{table}

 Table II summarizes the gate error budgeting of our protocol under different practical imperfections. Among them, although
 the dominant gate error remains 
 the position fluctuation, e.g. it stays at the level of $10^{-2}$ for the $q\neq 0$ case; yet this value has been greatly decreased by at least one order of magnitude as compared to that in the reported antiblockade-gate schemes \textcolor{black}{\cite{su2017fast,su2021dipole,wu2021one,shao2014one}}. Besides our protocol suffers a relatively larger sensitivity ($\sim 10^{-3}$) to the fluctuation of excitation lasers because of the pulse optimization. Other errors due to finite radiative lifetime of Rydberg states, the Doppler dephasing as well as the inhomogeneous Rabi frequency, are all contributing negligible ($10^{-5}\sim 10^{-4}$). After taking into account all error sources we can obtain a more conservative lower bound on the predicted gate fidelity $\mathcal{F}\geq 0.9143$. 
  The potential for a higher-fidelity antiblockade gate can depend on a further improvement in the tolerance to the position fluctuation, especially by using the cooling method. 
  

In conclusion we have shown the {\it modified} antiblockade effect cooperating with pulse optimization, can lead to Rydberg antiblockade SWAP gates with unprecedented tolerance, arising the experimental demonstration for them with great promise.
Through optimization we find the time spent in the double Rydberg state can be reduced by more than 70$\%$ as compared to that in the {\it perfect} antiblockade case, which does not depend on the realistic shape of the laser pulses. For this reason the insensitivity of such gates to the position fluctuations can be improved by one order of magnitude, arising a quite competitive gate fidelity of $\mathcal{F}\geq 0.9143$ in the presence of significant position deviation 
for a practical temperature $T\sim20$ $\mu$K. Larger gate tolerance maybe achievable by using more adequate cooling technique \cite{2015Mesoscopic}, optimal three-dimension geometric configuration \cite{PhysRevApplied.18.034072}, fault-tolerant nonadiabatic geometric quantum computation \cite{2022Error,Kang_2023} and other advanced optimization algorithm \textcolor{black}{\cite{PRXQuantum.4.020336}} et.al. \textcolor{black}{The prospect for realizing fast and high-fidelity state swapping with gate-duration-optimized pulses is presented in appendix B. We numerically confirm that an increase in the restriction of maximal Rabi frequency to be $2\pi\times 50$ MHz, can increase the conservative gate fidelity to 0.9550 within an optimal gate time $T_g=0.1177$ $\mu$s.} Finally, the fast and high-fidelity SWAP gate can be used as a tool for other important applications in the frontier of quantum information science, for example constructing scalable superconducting quantum processors \cite{PhysRevA.108.032615}.

\begin{acknowledgments}
We acknowledge financial support 
by the NSFC under Grants No. 12174106, No.11474094,  No.11104076, No.12274376, No.12304407, by the Science and Technology Commission of Shanghai Municipality under Grant No.18ZR1412800.
\end{acknowledgments}

\appendix

\section{Derivation of effective Hamiltonian Eqs.(4-6)} \label{aa}

In this appendix, we illustrate the derivation of the effective Hamiltonians in Eqs.(\ref{00e}-\ref{eff}). We recall the total Hamiltonian of the system, as 
\begin{equation}
\hat{\mathcal{H}}=\hat{\mathcal{H}}_{1}\otimes \hat{\mathcal{I}}_2 +\hat{\mathcal{I}}_1 \otimes\hat{\mathcal{H}}_{2} +\hat{V}_{rr} 
\end{equation}
with $\hat{\mathcal{I}}_j$ a $3\times 3$ identity matrix and $j=1,2$. Here the single-atom Hamiltonian in the Schr\"{o}dinger picture reads 
\begin{eqnarray}
    \hat{\mathcal{H}_j} &=& \sum_{i=0,1,r}\omega_i|i\rangle_j\langle i| + \Omega_1 |r\rangle_j\langle 0|e^{-i\omega_{l1}t} \nonumber\\
    &+& \Omega_2 |r\rangle_j\langle 1|e^{-i\omega_{l2}t} + \text{H.c.}
\end{eqnarray}
where $\omega_i$($i=0,1,r$) describes the frequency of the atomic level $|i\rangle$, and $\omega_{l1(l2)}$ represents the frequency of the optical driving fields. While turning to the interaction picture with respect to a rotating frame we have
\begin{equation}
\hat{\mathcal{H}_{j}}=\Omega_1e^{-i\Delta_1 t}|r\rangle_j\langle 0|+\Omega_2e^{-i\Delta_2 t}|r\rangle_j\langle 1|+\text{H.c}
\end{equation}
where we introduce the detuning parameters $\Delta_1 =\omega_{l1}-(\omega_{r}-\omega_{0})$, $\Delta_2=\omega_{l2}-(\omega_{r}-\omega_{1})$ for simplicity. Besides
the third term in (A1)
\begin{equation}
\hat{V}_{rr}=V|rr\rangle\langle rr|
\label{vrrn}
\end{equation}
is the Rydberg-mediated interaction as two atoms simultaneously occupy the Rydberg state and $V=C_6/r_0^6$ stands for the non-fluctuated {\it vdWs} interaction strength.


To simplify the calculation we first turn into the frame of two-atom basis states in which the above single-atom Hamiltonian $\hat{\mathcal{H}}_j$ could be re-expressed as
\begin{eqnarray}
\hat{\mathcal{H}}_{1}&=&\Omega _1e^{-i\Delta_1 t}(|r1\rangle\langle 01|+|r0\rangle\langle 00|+|rr\rangle\langle 0r|)\\
&+&\Omega _2e^{-i\Delta_2 t}(|r1\rangle\langle 11|+|r0\rangle\langle 10|+|rr\rangle\langle 1r|) + \text{H.c.} \nonumber\\
\hat{\mathcal{H}}_{2}&=&\Omega _1e^{-i\Delta_1 t}(|1r\rangle\langle 10|+|0r\rangle \langle 00|+|rr\rangle\langle r0|)  \\
&+&\Omega _2e^{-i\Delta_2 t}(|1r\rangle\langle 11|+|0r\rangle\langle 01|+|rr\rangle\langle r1|) +\text{H.c.} \nonumber
\end{eqnarray}
with which the total Hamiltonian (A1) of the system reduces to
\begin{equation}
\hat{\mathcal{H}} = \hat{\mathcal{H}}_1 + \hat{\mathcal{H}}_2 + \hat{V}_{rr} .
\label{a5}
\end{equation}
To demonstrate the role of the Rydberg-mediated interaction we rotate $\hat{\mathcal{H}}$ in (\ref{a5}) with respect to a unitary operator $\hat{U}=\text{exp}[{-i(\Delta_1+\Delta_2)t|rr\rangle\langle rr|}]$, which arises
\begin{equation}
\hat{\mathcal{H}}^\prime=\hat{U}^\dagger\hat{\mathcal{H}}\hat{U}+i\hbar\frac{\partial\hat{U}^\dagger}{\partial t}\hat{U}  =  \hat{\mathcal{H}}_0 + \hat{\mathcal{H}}_I
\end{equation}
where the products are given by
\begin{eqnarray}
 \hat{\mathcal{H}}_0 &=& \Omega _1e^{-i\Delta_1 t}(|r1\rangle\langle 01|+|r0\rangle\langle 00|)+\Omega _1e^{i\Delta_2 t}|rr\rangle \langle 0r| \nonumber \\
 &+& \Omega _2e^{-i\Delta_2 t}(|r1\rangle\langle11|+|r0\rangle\langle 10|)+\Omega _2e^{i\Delta_1 t}|rr\rangle\langle 1r| \nonumber\\
&+&\Omega _{2}e^{-i\Delta_2 t}({|0r\rangle\langle01|+|1r\rangle\langle 11|)+\Omega _{2}e^{i\Delta_1t}|rr\rangle \langle r1|} \nonumber\\
&+&\Omega _1e^{-i\Delta_1 t}{(|0r\rangle \langle 00|+|1r\rangle \langle 10|)+\Omega _1e^{i\Delta_2 t}|rr\rangle \langle r0|}) +\text{H.c}  \nonumber \\
 \hat{\mathcal{H}}_I &=& (V-\Delta_1-\Delta_2) |rr\rangle\langle rr|  \nonumber
\end{eqnarray}
\textcolor{black}{We aim at using the laser detuning to compensate the energy shift induced by the Rydberg-mediate interaction so it satisfies $V=\Delta_1+\Delta_2+q$ yielding $\hat{\mathcal{H}}_I = q|rr\rangle\langle rr|$, which ensures a nearly-resonant two-photon transition with respect to $|rr\rangle$. Clearly when $q=0$, it means the {\it perfect} antiblockade condition in which we can arrange both atoms to be excited to the Rydberg state. 
If $q$ is a nonzero value it represents the {\it modified} antiblockade which can modify the performance of some unwanted ac Stark shift and 
be effectively used for a SWAP gate task.}

Next, we rewrite the Hamiltonian $\hat{\mathcal{H}}_0$ as a more universal form 
\begin{equation}
\hat{\mathcal{H}}_0=\sum_{n=1,2}\hat{h}_n^\dagger e^{i\Delta_n t} + \hat{h}_n e^{-i\Delta_n t}  
\end{equation}
in which $\hat{h}_1$ and $\hat{h}_2$ take exact forms of
\begin{eqnarray}
\hat{h}_1&=& \Omega _1 (|r1\rangle\langle 01|+|r0\rangle\langle 00|+|0r\rangle \langle 00|+|1r\rangle \langle 10|) \nonumber\\
&+& \Omega _2(|1r\rangle\langle rr|+|r1\rangle\langle rr|) \nonumber\\
\hat{h}_2&=& \Omega _2 (|0r\rangle\langle 01|+|1r\rangle\langle 11|+|r1\rangle \langle 11|+|r0\rangle \langle 10|) \nonumber\\
&+& \Omega _1(|0r\rangle\langle rr|+|r0\rangle\langle rr|) .
\label{hh}
\end{eqnarray}
Furthermore, if the parameters satisfy $|\Delta_1| \gg \Omega_1^{\max}$, $|\Delta_2| \gg \Omega_2^{\max}$(large-detuning assumption) 
it is convenient to get a direct two-photon transition between $|rr\rangle$ and $|01\rangle$($|10\rangle$) with other singly-excited Rydberg states strongly suppressed for large detunings. Here we confine our interest to the dynamics which are time averaged over a longer period than any oscillation periods \cite{PhysRevA.82.052106}. So after neglecting all oscillation terms $\propto e^{i(\Delta_n\pm\Delta_m)t}$($n\neq m$) we finally arrive at a simple 
effective form for $\hat{\mathcal{H}}_0$ \cite{2010Quantum}
\begin{equation}
    \hat{\mathcal{H}}_{0,eff} = \frac{1}{\Delta_1} [\hat{h}_1^\dagger,\hat{h}_1]  +  \frac{1}{\Delta_2} [\hat{h}_2^\dagger,\hat{h}_2]
    \label{Hp}
\end{equation}
By inserting Eq.(\ref{hh}) into (\ref{Hp}) the effective Hamiltonian can be changed to:
\begin{widetext}
\begin{eqnarray}
  \hat{\mathcal{H}}_{0,eff} &=&  - (\frac{\Omega_{1}^2}{\Delta _{1}}+\frac{\Omega_{1}^2}{\Delta _{2}})|r0 \rangle\langle 0r| - (\frac{\Omega_{1}^2 }{\Delta _{1}} + 
   \frac{\Omega_{1}^2+ \Omega_{2}^2}{\Delta _{2}})(|r0 \rangle\langle r0| + |0r \rangle\langle 0r|)
  -(\frac{\Omega_{2}^2}{\Delta _{1}}+\frac{\Omega_{2}^2 }{\Delta _{2}})|r1 \rangle\langle 1r| \nonumber \\
  &-& (\frac{\Omega_{2}^2 }{\Delta_{2}}+\frac{\Omega_1^2+\Omega_{2}^2}{\Delta _{1}})(|r1 \rangle\langle r1|+ |1r \rangle\langle 1r|)
   +  (\frac{\Omega_{1}\Omega_{2} }{\Delta _{1}}+\frac{\Omega_{1}\Omega_{2} }{\Delta _{2}})(|rr \rangle\langle 01|+ |rr \rangle\langle 10| ) \label{eff00} \\
   &+&(\frac{\Omega_{1}^2 }{\Delta _{1}}+\frac{\Omega_{2}^2 }{\Delta _{2}})(|01 \rangle\langle 01|+|10 \rangle\langle 10|)
  +  \frac{2\Omega_{1}^2}{\Delta _{1}} |00 \rangle\langle 00|+\frac{2\Omega_{2}^2 }{\Delta _{2}}|11 \rangle\langle 11|+(\frac{2\Omega_{2}^2 }{\Delta _{1}}+
   \frac{2\Omega_{1}^2 }{\Delta _{2}})|rr \rangle\langle rr| +\text{H.c.} \nonumber
\end{eqnarray}  
\end{widetext}
Via following Eq.(\ref{eff00}) we observe that the singly-excited states such as $\{|0r\rangle,|r0\rangle,|1r\rangle,|r1\rangle\}$ are now in a closed subspace and decoupled to the initial states $\{|00\rangle,|01\rangle,|10\rangle,|11\rangle\}$ due to the assumption of large laser detunings. So it is safe to drop these trivial terms and the total Hamiltonian $\hat{\mathcal{H}}^\prime$ can be effectively described by
\begin{eqnarray}
  \hat{\mathcal{H}}^\prime_{eff} &=& 
  \hat{\mathcal{H}}^\prime_{0,eff} + \hat{\mathcal{H}}_I \nonumber\\
  &=&(\frac{\Omega_{1}\Omega_{2} }{\Delta _{1}}+\frac{\Omega_{1}\Omega_{2} }{\Delta _{2}})(|rr \rangle\langle 01|+ |rr \rangle\langle 10| )+\text{H.c.} \nonumber\\
  &+&(\frac{\Omega_{1}^2 }{\Delta _{1}}+\frac{\Omega_{2}^2 }{\Delta _{2}})(|01 \rangle\langle 01|+|10 \rangle\langle 10| ) +  \frac{2\Omega_{1}^2}{\Delta _{1}} |00 \rangle\langle 00| \nonumber\\
  &+&\frac{2\Omega_{2}^2 }{\Delta _{2}}|11 \rangle\langle 11|+(\frac{2\Omega_{2}^2 }{\Delta _{1}}+
   \frac{2\Omega_{1}^2 }{\Delta _{2}}+q)|rr \rangle\langle rr|     
   \label{effHo}
\end{eqnarray}

Based on Eq.(\ref{effHo}) one can see the collective ground state $|00\rangle$ or $|11\rangle$ stays decoupled arising
\begin{equation}
  \hat{\mathcal{H}}_{00,eff} =    \frac{2\Omega_{1}^2}{\Delta _{1}} |00 \rangle\langle 00| \text{  or  }
  \hat{\mathcal{H}}_{11,eff} = \frac{2\Omega_{2}^2 }{\Delta _{2}}|11 \rangle\langle 11|
 \end{equation}
for the initialization of states $|00 \rangle$ and $|11 \rangle$. However, when the initial state is $|01\rangle$(equivalent to $|10\rangle$) one has
 \begin{eqnarray}
  \hat{\mathcal{H}}_{01,eff} &=&  (\frac{\Omega_{1}\Omega_{2} }{\Delta _{1}}+\frac{\Omega_{1}\Omega_{2} }{\Delta _{2}})(|rr \rangle\langle 01|+|01 \rangle\langle rr|) +\text{H.c.}
   \nonumber \\
   &+&(\frac{2\Omega_{2}^2 }{\Delta _{1}}+
   \frac{2\Omega_{1}^2 }{\Delta _{2}}+q)|rr \rangle\langle rr|  \nonumber  \\
   &+&(\frac{\Omega_{1}^2 }{\Delta _{1}}+\frac{\Omega_{2}^2 }{\Delta _{2}})(|01 \rangle\langle 01|+|10 \rangle\langle 10| )
   \label{effH}
\end{eqnarray}
It is worthwhile to stress that although
 the $\hat{\mathcal{H}}_{01,eff}$ involves the required swapping dynamics between two target states $|01\rangle$ and $|10\rangle$, yet which is clearly mediated by the double Rydberg state $|rr\rangle$ because of the antiblockade facilitation,
 increasing its sensitivity to the position error. Luckily the presence of factor $q(\neq 0)$ can modify the Stark shift with respect to $|rr\rangle$ suppressing the population on it by using an off-resonance detuning. This makes the resulting gate insusceptible to the interaction fluctuation. In addition the ac Stark shift to the ground states $|01\rangle$ and $|10\rangle$(the last term in Eq.(\ref{effH})) would no doubt influence the swapping dynamics. Applying auxiliary atom-field couplings to overcome it as done by  \textcolor{black}{\cite{su2016one}}, may increase the experimental complexity. We identify that, 
 the way of pulse optimization in our protocol can minimize this influence without auxiliary fields, promising for a high-fidelity SWAP gate with improved robustness. \textcolor{black}{Finally, note that our protocol benefiting from the combination of the {\it modified} antiblockade effect and the pulse optimization, is promising for other two-qubit quantum gates with a different pulse design. For example, a CNOT gate requires an asymmetric pulse driving in which atom 1 is only driven by $\Omega_2(t)$ for the $|1\rangle\to|r\rangle$ transition leaving $|0\rangle$ idle. In this case the antiblockade facilitation could allow for an effective coupling between $|10\rangle$ and $|11\rangle$ achieving a robust two-qubit CNOT gate.}


\section{Achieving fast state swapping with pulse optimization} \label{ab}

\begin{figure*}
\centering
\includegraphics[width=13.8cm,height=9.0cm]{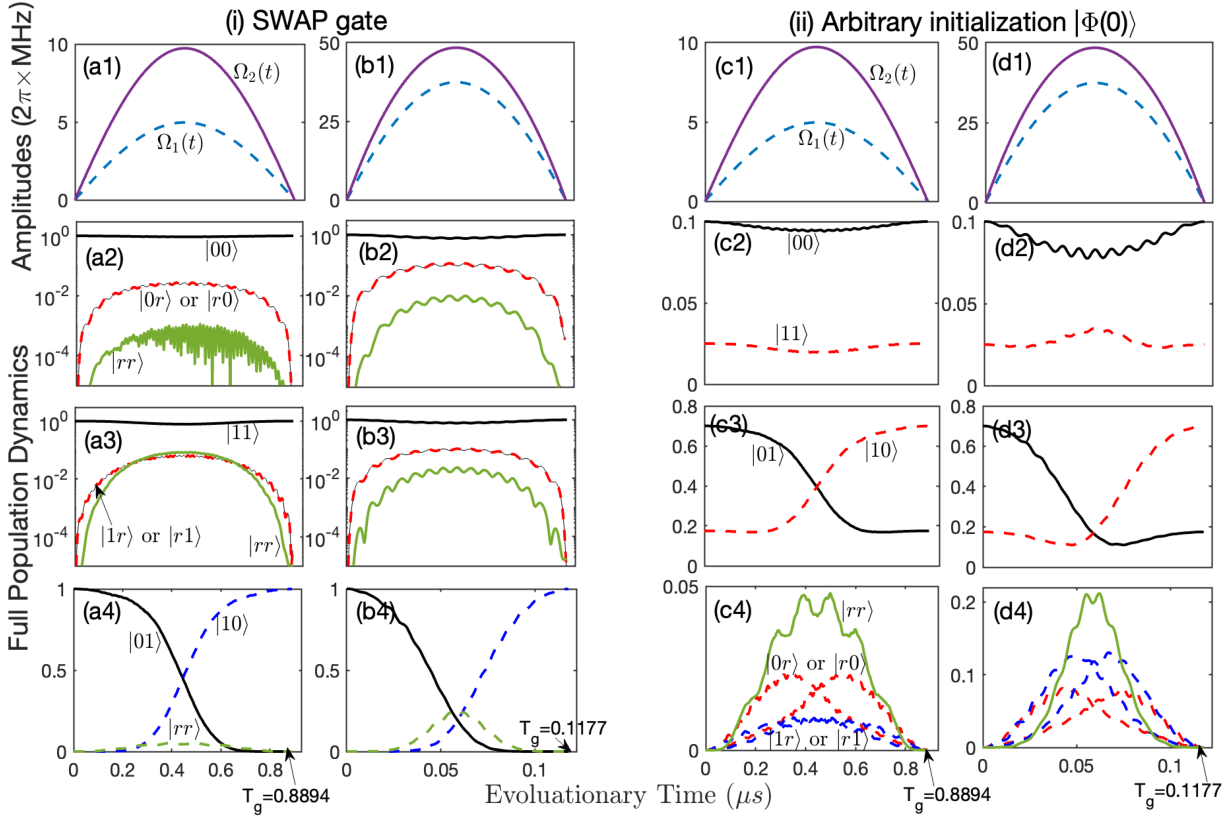}
\caption{Optimal Gaussian pulse shapes and the full population dynamics based on (i) the SWAP gate protocol and (ii) an arbitrary initialization $|\Phi(0) =\frac{1}{\sqrt{10}}|00\rangle + \sqrt{\frac{7}{10}}|01\rangle + \sqrt{\frac{7}{40}}|10\rangle + \frac{1}{\sqrt{40}}|11\rangle $. By constraining the maximal Rabi frequencies $(\Omega_1^{\max},\Omega_2^{\max})$ to be $2\pi\times10$ MHz for columns (a,c) and $2\pi\times50$ MHz for columns (b,d), the optimal gate durations are $T_g=0.8894$ $ \mu$s and $0.1177$ $\mu$s, correspondingly.
 All linetypes are denoted in (a1-a4) and (c1-c4).}
\label{Fig5:super}
\end{figure*}

In this appendix we provide extended calculations for a fast swapping operation in practice. Inspired by the methods used in \cite{Jandura2022timeoptimaltwothree} we find that, when the gate duration $T_g$ is also varied serving as another ancillary optimal parameter the SWAP gate performance can further be improved with a higher tolerance. 
To show the role of $T_g$ optimization we focus on two cases: one with a restriction of $(\Omega_1^{\max},\Omega_2^{\max})/2\pi\leq 10$ MHz and a second with $(\Omega_1^{\max},\Omega_2^{\max})/2\pi\leq 50$ MHz. Such choices will lead to an approximate timescale level for the gate duration which are $1.0$ $\mu$s and $0.1$ $\mu$s for maintaining complete circles of rotation. Note that when we vary the pulse parameters, a simultaneous optimization for the gate duration can also minimize the time spent in the Rydberg state.

In this extended calculation all parameters including $T_g$ are tunable so as to maximize the perfect gate fidelity in the absence of any technical noise. To be specific we also extend the simplified model in Fig.\ref{Fig1:model} by using different branching ratios with the leakage state $|\alpha\rangle$
which are $\eta_{r\to1} = 1/8$, $\eta_{r\to2} = 1/8$, $\eta_{r\to\alpha} = 3/4$. State $|\alpha\rangle$ represents all ground magnetic sublevels $|5S_{1/2},F=1,m_F = \pm1\rangle$ and $|5S_{1/2},F=2,m_F =\pm 1,\pm 2 \rangle$ except $|0\rangle$, $|1\rangle$.
In this practical situation we find the success of the optimization does not affected by the increase of parameters to be optimized. Finally we obtain two sets of new parameters for the gate
\begin{eqnarray}
\text{(1)}\text{   }   (q,\Delta_1,\Delta_2)/2\pi &=& (6.051,22.951,41.488)  \text{ MHz} \nonumber \\
(\Omega_1^{\max},\Omega_2^{\max})/2\pi &=&  (4.969,9.728) \text{ MHz} \nonumber \\
      (\omega_1,\omega_2) &=& (0.4948,0.5318) \text{ }\mu\text{s} \nonumber\\ 
      T_{g}&=&0.8894 \mu s \label{ne1} 
\end{eqnarray}

\begin{eqnarray}
(2)\text{   }   (q,\Delta_1,\Delta_2)/2\pi &=& (33.429,-94.133,131.194)  \text{ MHz}  \nonumber\\  
(\Omega_1^{\max},\Omega_2^{\max})/2\pi &=&  (37.413,48.404) \text{ MHz} \nonumber \\
      (\omega_1,\omega_2) &=& (0.0586,1.6082) \text{ }\mu\text{s} \nonumber\\ 
      T_{g}&=&0.1177 \mu s \label{ne2}
\end{eqnarray}
respectively corresponding to the restrictions $\Omega_{1(2)}^{\max}/2\pi\leq 10$ MHz and $50$ MHz set for the numerical optimization.

\begin{table*}
\tiny
\caption{\label{tab:table2} Error budget for the two-qubit SWAP gate and for an arbitrary initialization $|\Phi(0)\rangle$. From left to right the numerical simulations are performed at different branching ratios or different $T_g$ values, corresponding to the cases graphically plotted in Fig.\ref{Fig2:dyna}(a5-e5), Fig.\ref{Fig5:super}(a1-a4), (b1-b4), (c1-c4) and (d1-d4), respectively. Finally the perfect fidelity is only contributed by the Rydberg decay error. Besides, a more conservative estimation for the state swapping fidelity is given accounting for the sum of all intrinsic and technical error sources.}
\renewcommand{\arraystretch}{1.2}
\setlength{\tabcolsep}{0.7mm}
  \centering
 \resizebox{\textwidth}{30mm}{
   \begin{tabular}{c c c c c  c c}
        \hline
        \hline
    Error Sources & \multicolumn{3}{c}{Error Budget(SWAP)} & \multicolumn{2}{c}{Error Budget($|\Phi(0)\rangle$)} \\
        \hline
         Branching ratios $(\eta_{r\to 1},\eta_{r\to 2},\eta_{r\to \alpha})$ &  ($\frac{1}{2},\frac{1}{2},0$) & ($\frac{1}{8},\frac{1}{8},\frac{3}{4}$) & ($\frac{1}{8},\frac{1}{8},\frac{3}{4}$) & ($\frac{1}{8},\frac{1}{8},\frac{3}{4}$) & ($\frac{1}{8},\frac{1}{8},\frac{3}{4}$)   \\
        \hline
         $q/2\pi$ (\text{MHz}) &  5.037  & 6.051 & 33.429 & 6.051 & 33.429 \\
         \hline
         $T_g$ ($\mu$s) &  1.0  & 0.8894  & 0.1177 & 0.8894 &  0.1177\\
        \hline
        Rydberg decay  & $3.18\times10^{-4}$ & $5.16\times10^{-4}$ & $4.58\times10^{-4}$ & $4.97\times10^{-4}$ &$5.83\times10^{-4}$\\
       \hline 
  Position fluctuation~($\sigma_{r}=47.34$ nm) &  $7.52\times 10^{-2}$ & $6.98\times 10^{-2}$ & $1.15\times 10^{-2}$ & $2.78\times 10^{-2}$ & $1.01\times 10^{-3}$\\
  Doppler dephasing~($T=20$ $\mu$K) &$3.80\times10^{-5}$ & $1.58\times10^{-5}$ & $1.22\times10^{-5}$ & $8.00\times10^{-6}$& $2.76\times10^{-5}$\\
  Inhomogeneous Rabi frequency~($\omega=2$ $\mu$m) &$5.02\times10^{-5}$ &$1.24\times10^{-5}$ & $2.76\times10^{-5}$ & $3.00\times10^{-6}$ &$2.74\times10^{-6}$\\
   \hline
  Laser intensity~($\delta_{\Omega}=5.0$ $\%$) &
  $3.37\times10^{-3}$ & $3.11\times10^{-3}$ & $3.11\times10^{-2}$ & $2.47\times10^{-4}$ & $1.73\times10^{-2}$\\
Laser phase~($\gamma_0/2\pi=100$ kHz)&
  $6.68\times10^{-3}$ & $6.82\times10^{-3}$ & $1.94\times10^{-3}$ & $5.23\times10^{-3}$ & $2.07\times10^{-3}$\\
        \hline
        \hline   
        State swapping fidelity (perfect) & 0.9997  & 0.9995 & 0.9995 & 0.9995 & 0.9994  \\       
        State swapping fidelity (conservative) & 0.9140  & 0.9200 & 0.9550 & 0.9662 & 0.9790  \\
         \hline
        \hline          
\end{tabular}}
\end{table*}

In the columns (a,b) of Fig.\ref{Fig5:super} we show the optimal pulse shapes $(\Omega_1(t),\Omega_2(t))$ as well as the corresponding population dynamics. As compared to the case (a5-e5) in Fig.\ref{Fig2:dyna}, we find that, together with the optimization for $T_g$ the time spent on $|rr\rangle$ could be slightly shortened $T_{rr}\approx 0.0216$ $\mu$s owing to the use of a small and optimal $T_g=0.8894$ $\mu$s($<1.0$ $\mu$s). The perfect gate fidelity stays at a high level $\mathcal{F}\approx 0.999484$, although this value is slightly smaller than the observed one $\mathcal{F}\approx 0.999682$ in {\it Case ii}($q\neq0$, Table I) because the population loss caused by the leakage state $|\alpha\rangle$ is irreversible.
In addition note that if the restriction for $\Omega_{1(2)}^{\max}$ is increased to $2\pi\times 50$ MHz, the $T_g$ required can be lowered to 0.1177 $\mu$s excellently agreeing with the relationship between $\Omega_{1(2)}^{\max}$ and $T_g$. From the error budget in Table III, it is apparent that the Rydberg decay representing the dominant source of error for the perfect gate fidelity, can not be easily affected by shortening the gate duration. Therefore the state swapping fidelity in the perfect case is almost the same. However, for a shorten gate duration at the expense of larger Rabi frequency the time spent $T_{rr}$ is able to decrease by one order of magnitude ($T_{rr}\approx 0.0085$ $\mu$s) which can no doubt make the gate more susceptible to other technical imperfections. As shown in Table III the gate tolerance to the position fluctuation obtains an explicit improvement($6.98\times 10^{-2}\to 1.15\times 10^{-2}$) owing to the suppression of the Rydberg-state duration. The only worse effect comes from the laser intensity fluctuation because it increases with the values of $\Omega_1^{\max}$ and $\Omega_2^{\max}$. For a fast state swapping operation the laser intensity noise will play an equivalently important role as the position fluctuation.
By taking account of different technical imperfections limiting the gate fidelity we finally demonstrate that a conservative fidelity for the fast SWAP protocol above 0.9550 is achievable.

In order to show the SWAP gate that can be used for an arbitrary state swapping we introduce a normalized two-qubit state serving as the initial state
\begin{equation}
    |\Psi(0)\rangle = A_{00}|00\rangle + A_{01}|01\rangle + A_{10}|10\rangle + A_{11}|11\rangle \label{idea}
\end{equation}
with $\sum_{i,j} |A_{i,j}|^2 = 1$ and $i,j\in\{0,1\}$. Such arbitrariness of the initialization can be used to study the fast state swapping of the scheme. Ideally for an arbitrary initialization $|\Phi(0)\rangle$ the output state should be $|\Psi(T_g)\rangle = A_{00}|00\rangle + A_{10}|01\rangle + A_{01}|10\rangle + A_{11}|11\rangle$. Here we modify the definition of average fidelity in Eq.(\ref{fde}) by
\begin{equation}
    \mathcal{F} = \text{Tr}\sqrt{\sqrt{U}\rho(t=T_g)\sqrt{U}}
\end{equation}
because the initial state $|\Phi(0)\rangle$ is individually normalized. $\rho$ is the practical density matrix at measured time $t=T_g$ and $U=|\Phi(0)\rangle\langle\Phi(0)|$ is the ideal swapping matrix. In the columns (c,d) of Fig.\ref{Fig5:super} we comparably show the time-dependent evolution of different population $|A_{i,j}(t)|^2$ for state $|ij\rangle$ by adopting the optimization parameters in (\ref{ne1}) and (\ref{ne2}). A complete error budget is presented in the last two columns of Table III. In general a fast and high-fidelity state swapping is achievable for an arbitrary initialization. The swapping fidelity considered for a non-fluctuated environment(perfect case), can sustain at a high value 0.9995(4) owing to the success of pulse optimization.
Whereas, as taking account of various technical limitations from a typical Rydberg experimental setup, we find that the position fluctuation and the laser intensity noise are still two dominant contributions, as same as in the SWAP gate protocol. But the greatest influence from a fluctuated position has been lowered to $1.01\times 10^{-3}$ because of the relatively small population in two swapping states $|01\rangle$ and $|10\rangle$.
Oppositely the laser intensity noise plays the most important role because the maximal Rabi frequency to be optimized has reached $2\pi\times 48.404$ MHz for $\Omega_2^{\max}$.
After considering the technical noises a conservative estimation for the predicted fidelity of state swapping can achieve 0.9662 and 0.9790. It is worthwhile to note that, 
achieving fast quantum state operations within a submicrosecond duration has promising advantages deserving more theoretical and experimental efforts in the future.

\end{document}